\documentclass[aip,pop,reprint,amsmath,amssymb,floatfix]{revtex4-2}

\usepackage{graphicx}			
\usepackage{subfig}				
\usepackage{epstopdf}			
\usepackage{bm}				
\usepackage{units}				
\usepackage{dcolumn}			
\usepackage{color}				
\usepackage[colorlinks=true, linkcolor=blue, citecolor=blue]{hyperref}	
\usepackage[font=footnotesize]{caption}
\newcommand\todo[1]{\textcolor{black}{#1}}

\begin{document}

\title{Laser-Driven, Ion-Scale Magnetospheres in Laboratory Plasmas. I. Experimental Platform and First Results}

\author{D. B. Schaeffer}
	\email{dereks@princeton.edu}
   	\affiliation{Department of Astrophysical Sciences,  Princeton University, Princeton, NJ 08540, USA}
\author{F. D. Cruz}
   	\affiliation{GoLP/Instituto de Plasmas e Fus\~{a}o Nuclear, Instituto Superior T\'{e}cnico, Universidade de Lisboa, 1049-001 Lisboa, Portugal}
\author{R. S. Dorst}
   	\affiliation{Department of Physics and Astronomy, University of California -- Los Angeles, Los Angeles, CA 90095, USA}
\author{F. Cruz}
   	\affiliation{GoLP/Instituto de Plasmas e Fus\~{a}o Nuclear, Instituto Superior T\'{e}cnico, Universidade de Lisboa, 1049-001 Lisboa, Portugal}
\author{P. V. Heuer}
   	\affiliation{Department of Physics and Astronomy, University of California -- Los Angeles, Los Angeles, CA 90095, USA}
\author{C. G. Constantin}
   	\affiliation{Department of Physics and Astronomy, University of California -- Los Angeles, Los Angeles, CA 90095, USA}
\author{P. Pribyl}
   	\affiliation{Department of Physics and Astronomy, University of California -- Los Angeles, Los Angeles, CA 90095, USA}
\author{C. Niemann}
   	\affiliation{Department of Physics and Astronomy, University of California -- Los Angeles, Los Angeles, CA 90095, USA}			
\author{L. O. Silva}
	\affiliation{GoLP/Instituto de Plasmas e Fus\~{a}o Nuclear, Instituto Superior T\'{e}cnico, Universidade de Lisboa, 1049-001 Lisboa, Portugal}
\author{A. Bhattacharjee}
   	\affiliation{Department of Astrophysical Sciences,  Princeton University, Princeton, NJ 08540, USA}
	\affiliation{Princeton Plasma Physics Laboratory, Princeton, NJ 08543, USA}


\begin{abstract}

Magnetospheres are a ubiquitous feature of magnetized bodies embedded in a plasma flow. While large planetary magnetospheres have been studied for decades by spacecraft, ion-scale ``mini'' magnetospheres can provide a unique environment to study kinetic-scale, collisionless plasma physics in the laboratory to help validate models of larger systems.  In this work, we present preliminary experiments of ion-scale magnetospheres performed on a unique high-repetition-rate platform developed for the Large Plasma Device (LAPD) at UCLA.  The experiments utilize a high-repetition-rate laser to drive a fast plasma flow into a pulsed dipole magnetic field embedded in a uniform magnetized background plasma. 2D maps of magnetic field with high spatial and temporal resolution are measured with magnetic flux probes to examine the evolution of magnetosphere and current density structures for a range of dipole and upstream parameters. The results are further compared to 2D PIC simulations to identify key observational signatures of the kinetic-scale structures and dynamics of the laser-driven plasma.  We find that distinct 2D kinetic-scale magnetopause and diamagnetic current structures are formed at higher dipole moments, and their locations are consistent with predictions based on pressure balances and energy conservation.

\end{abstract}

\maketitle


\section{Introduction}

Magnetospheres form when a plasma flow impacts a magnetic obstacle, such as the interaction between the solar wind and planets with intrinsic magnetic fields in the heliosphere.  The plasma flow is largely stopped at the magnetopause, where the kinetic ram pressure of the flow balances the magnetic field pressure, and moves around the obstacle to form a magnetotail downstream.  If the incoming flow is super-Alfv\'{e}nic, a bow shock can also be created ahead of the magnetopause, leading to the generation of a magnetosheath composed of shocked plasma.  Additionally, if the magnetic obstacle is embedded in a background magnetic field (analogous to the interplanetary magnetic field [IMF]), the orientation of the obstacle relative to the background field can have significant effects on the global magnetic structure, including magnetic reconnection. These features are readily observed at planets, including the Earth, which has been studied \textit{in situ} by spacecraft for decades \cite{pulkkinen2007space,eastwood2008science,russell2016,Borovsky2018}.

To first order, the magnetic obstacles of interest can be modeled as dipoles, so that magnetospheres can be characterized by the so-called Hall parameter $D=L_M/d_i$, where $L_M$ is the distance from the dipole center to the magnetopause, and $d_i=c/\omega_{pi}$ is the upstream ion inertial length.  \todo{In other words, $D$ can be interpreted as the effective size of the magnetic obstacle \cite{omidi_dipolar_2004,shaikhislamov_mini-magnetosphere:_2013}.}  Planetary magnetospheres are large; indeed, for Earth $D>600$.  If the magnetopause distance is comparable to the ion inertial length, though, ion-scale magnetospheres can form.  These mini-magnetospheres have been observed \todo{in a variety of natural systems, including around comets \cite{nilsson_birth_2015} and locally magnetized regions on the Moon \cite{halekas_density_2008,wieser_first_2010,lue_strong_2011,bamford_minimagnetospheres_2012,halekas_evidence_2014},} and are of interest for spacecraft propulsion \cite{moritaka_momentum_2012}.  However, understanding both their local and global scale structures (both kinetic and system size) has been constrained by available spacecraft diagnostics and single-spacecraft trajectories.  \todo{These limitations have been partially addressed by numerical efforts, where fully-kinetic \cite{karimabadi_the-link_2014,gonzalez_magnetic_2016} and hybrid-fluid-kinetic simulations \cite{winske_hybrid_2003,lin_three-dimensional_2005,blanco-cano_global_2009,omelchenko_hypers_2021} have shown the importance of expanding beyond MHD descriptions when modeling magnetospheres, including mini-magnetospheres \cite{vernisse_stellar_2017,deca_electron_2017,deca_building_2019}.}

Laboratory experiments can thus help address key questions about ion-scale magnetospheres and complement spacecraft and numerical efforts by providing controlled and reproducible conditions and measurements of both global and kinetic scales.  2D hybrid simulations (kinetic ions, fluid electrons)  \cite{omidi_hybrid_2002,omidi_dipolar_2004,omidi_macrostructure_2005,gargate_hybrid_2008} have shown that different regimes of magnetosphere formation can be parameterized with $D$.  The results indicate that for $D\ll 1$, there is no appreciable flow deflection, though whistler waves can develop in the obstacle's wake.  At larger $D\sim1$, there is some pile-up of plasma at the magnetopause, resulting in a fast mode bow wave and some heating in the magnetotail.  Only in the large-scale Hall regime ($D>20$) are fully formed magnetospheres, including the presence of a bow shock, observed.  More recently, 3D fully kinetic particle-in-cell (PIC) simulations have shown that bow shocks can form when $L_M/\rho_i>1$, where $\rho_i$ is the upstream ion gyroradius \cite{cruz_formation_2017}.  This condition is equivalent to $D>M_A$, where $M_A$ is the Alfv\'{e}nic Mach number of the plasma flow and $\rho_i=M_A d_i$.  These simulations thus predict that for low Mach number flows, the conditions necessary to form a magnetosphere are less stringent than those suggested by the earlier hybrid simulations.

Following the development of these large-scale magnetospheric simulations, there has been increased interest in laboratory experiments over the past couple decades.  Work by Yur \textit{et al.} \cite{yur_laboratory_1995,yur_magnetotail_1999} used a plasma gun to study the structure of the magnetotail and its dependence on the orientation of a background magnetic field.  Utilizing a super-Alfv\'{e}nic plasma flow and magnetic dipole, early experiments by Brady \textit{et al.} \cite{brady_laboratory_2009} confirmed that the location of the flow-dipole pressure balance ($L_M$) could be modeled with MHD in the small Hall regime ($D<<1$).  Bamford \textit{et al.} \cite{bamford_minimagnetospheres_2012} used a plasma wind tunnel to study similarly weak interactions relevant to lunar mini-magnetospheres.  Experiments by Zakharov \textit{et al.} \cite{zakharov_laser-plasma_2009}, and later Shaikhislamov \textit{et al.} \cite{shaikhislamov_mini-magnetosphere:_2013,shaikhislamov_laboratory_2014}, utilized a high-energy laser to drive a super-Alfv\'{e}nic plasma flow into a magnetic dipole, and in several cases incorporated a theta pinch to provide an ambient plasma and external magnetic field.  While these experiments achieved $D\sim1<M_A$, measurements were limited to 1D magnetic field and plasma density profiles.

To overcome these limitations, we have developed a new experimental platform to study ion-scale magnetospheres on the Large Plasma Device (LAPD) at UCLA.  This platform uniquely combines the large-scale, ambient magnetized plasma provided by the LAPD, a fast collisionless plasma flow generated by a laser driver, and a rotatable pulsed dipole magnetic field, all operating at high-repetition-rate ($\sim1$ Hz).  Utilizing motorized probes, we can measure for the first time the 3D structure of mini-magnetospheres over a wide range of parameters and magnetic geometries.  \todo{The goals of these experiments are 1) to study the formation and structure of laser-driven ion-scale magnetospheres, 2) to study the effect of magnetic reconnection on magnetosphere dynamics, and 3) to utilize super-Alfv\'{e}nic flows to generate and study bow shocks in the $D>M_A>1$ regime.}

In this paper, we report the first results from experiments on laser-driven, ion-scale magnetospheres on the LAPD \todo{that focus on the formation of magnetosphere structure with sub-Alfv\'{e}nic flows.}  In the experiments, a laser-driven plasma expands supersonically into a dipole magnetic field embedded in an ambient magnetized plasma, so that the total magnetic field topology is analogous to that of the Earth's magnetosphere superposed with a northward IMF.  By measuring 2D planes of the magnetic field over thousands of shots, we demonstrate the formation of a magnetopause and show how its structure evolves in time for a range of dipole strengths in the $D\sim1$ regime.  The results are consistent with 2D PIC simulations modeled after the experiments, which show that both the ambient and laser-produced ions play a key role in the formation of the magnetosphere.  Additional simulation results are presented as the second part of this series \cite{cruz_laser-driven_2021}, hereafter referred to as Part II.

The paper is organized as follows.  Section \ref{sec:setup} describes the setup of the experiments and typical parameters.  Section \ref{sec:results} discusses the main results, including: the performance of the dipole magnet, fast-gate images of the laser-driven plasma, measurements of the magnetosphere, and comparisons with simulations.  The interpretation of the results are discussed in Sec.~\ref{sec:discuss} before concluding in Sec.~\ref{sec:conc}.

\section{Experimental Setup}\label{sec:setup}

\begin{figure}[t]
	\centering
	\includegraphics{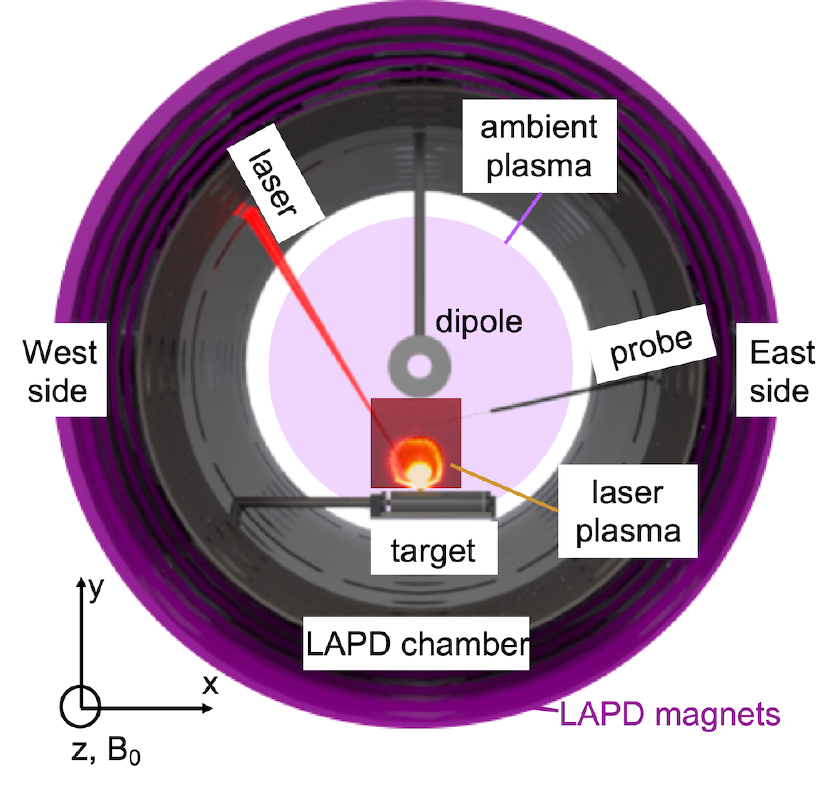}
	\caption{Schematic of the experimental setup on the LAPD.  A laser ablates a plastic target to create a supersonic plasma flow, which flows towards a dipole magnet inserted into the LAPD from the top.  The dipole magnet is embedded in a uniform magnetized background plasma generated by the LAPD.  Probes inserted from the east port collect volumetric data from the regions around the dipole.  A fast-gate image shows the expansion of the laser-driven plasma.}
	\label{fig:setup}
\end{figure}

The experiments were carried out on the Large Plasma Device (LAPD) at UCLA, operated by the Basic Plasma Science Facility (BaPSF), and combined a magnetized ambient plasma, a fast laser-driven plasma flow, and a current-driven dipole magnet.  A schematic of the experimental setup is shown in Fig. \ref{fig:setup}, and typical background and laser-driven plasma parameters are listed in Table \ref{tab:params}.  The LAPD \cite{gekelman_design_1991} is a cylindrical vacuum vessel (20 m long by 1 m diameter) that can generate a steady-state ($\sim$15 ms), large volume ($>50$ cm across the plasma column), magnetized ambient plasma at high repetition (up to 1 Hz).  The machine can produce variable background magnetic fields (200-1500 G), variable ambient gas fills (e.g. H, He), and variable ambient densities ($10^{11}-10^{13}$ cm$^{-3}$).  The ambient plasma is generated from the combination of two cathodes.  A BaO-coated Ni cathode generates a \O60 cm, lower-density ($n\sim2\times10^{12}$ cm$^{-3}$) main plasma, while a LaB$_{6}$ (lanthanum hexaboride) cathode generates a smaller \O20 cm, higher-density ($n\sim2\times10^{13}$ cm$^{-3}$) core plasma roughly centered on the main one.  The ambient plasma has a typical electron temperature $T_{e}\approx5$-$10$ eV and ion temperature $T_{i}\approx1$ eV.  The background field is oriented axially ($\hat{z}$) along the machine, with $\hat{x}$ oriented horizontally perpendicular to the field and $\hat{y}$ oriented vertically.

The supersonic plasma flow was generated by the high-repetition-rate Peening \cite{dane_design_1995, schaeffer_a-platform_2018} laser, operated by the UCLA High Energy Density Plasma (HEDP) group \cite{constantin_collisionless_2009}.  The Peening laser (1053 nm) can deliver energies up to 20 J with a pulse width of 15 ns (FWHM), yielding typical intensities of $10^{12}$ W/cm$^{2}$ and repetition rates up to 4 Hz.  The output laser energy, pulse shape, diffraction-limited focus, and beam pointing are stable to within within 5\% \cite{schaeffer_a-platform_2018}.

The dipole magnet consisted of an epoxy-covered 24-turn copper coil with integrated water cooling and a non-magnetic stainless steel housing and support shaft. It has a 14 cm outer diameter and a 4 cm inner diameter.  A pulsed power cabinet was capable of driving up to 7 kA at 800 V through the coil, corresponding to peak on-axis magnetic fields of 15 kG (magnetic moment $M\approx2.2$~kAm$^2$, sufficient to achieve large standoff values), which is approximately constant for several tens of $\mu$s (i.e. the whole experiment).  The field could be pulsed up to 1 Hz ($1/4$ Hz at the highest currents), and the water cooling allows the magnet to remain at room temperature throughout operation.

The target was a long, 5 cm diameter cylindrical rod of high-density polyethylene (C$_2$H$_4$) plastic.  The target was mounted on a 2D stepper motor drive synchronized with the laser, which translated and rotated the target in a helical pattern.  Each target position was repeated three times and then moved to provide a fresh surface.  A single target could thus be used for up to $2\times10^{4}$ laser shots. 

The dipole magnet was inserted from the top flange, so that the distance from the target to the dipole was variable, with the dipole orientation such that the dipole axis was along $z$ and rotatable about the $y$ axis. The lasers were timed to fire at the peak of the dipole field (time $t_0$), and the experiment lasted for a few tens $\mu$s, well within the long ($\sim10$ ms) lifetime of the ambient plasma.  The target and probes were set up in a ``dayside'' configuration, \todo{analogous to the sun-facing region of Earth's magnetosphere, as follows}.  The target was inserted through the bottom 45$^{\circ}$ west-side port at an angle parallel to the bottom flange (i.e. along $\hat{x}$), which placed the target surface 27.5 cm from the chamber center.  The laser was routed from the laserbay, though the LAPD room ceiling, to the top 45$^{\circ}$ west-side port, where it was focused and sent through a vacuum window, impinging the target at an angle of 30$^{\circ}$ relative to the target surface normal.  The resulting laser plasma expanded up towards the dipole, and probes were inserted from the east-side.  This arrangement allowed probes to move throughout the dayside region of the magnetosphere.  The laser, target, and pulsed dipole magnet were synchronized to the LAPD, and they all operated at a repetition rate of $1/4$ Hz to allow time for the diagnostics to position themselves between shots.

During the experiments, the ambient gas fill was H and the background magnetic field was set to 300 G.  The dipole magnet was arranged such that the dipole magnetic field was parallel to the background field in the dayside region.  The laser ablated a highly-energetic supersonic plasma, consisting of both C and H ions from the target, that expanded towards the dipole and transverse to the background (LAPD) magnetic field.  The interaction between the flowing and stationary ions is highly collisionless (mean free path $\gg$ system size) due to the high flow speeds.  The background electrons were also collisionless as the electron-ion collision time was much larger than the electron gyroperiod $\omega_{ce0}\tau_{ei}\approx500$.


The magnetic field topology and dynamics were measured with 3 mm diameter, 3-axis 10-turn magnetic flux (``bdot'') probes \cite{everson_design_2009}.  The probe signals were passed through a 150 MHz differential amplifier and coupled to either fast (1.25 GHz) or slow (100 MHz) 10-bit digitizers, and then numerically integrated to yield magnetic field amplitude.  To acquire data, the probes were positioned by a 3D motorized probe drive (resolution $<0.1$ cm) \cite{gekelman_the-upgraded_2016} in between shots.  Datasets were compiled by moving the probes in small increments of 0.25 cm with 3 shots per position for statistics.

Fast-gate ($\sim10$ ns) imaging \cite{heuer_fast_2017} was used to acquire 2D snapshots of plasma self-emission during the interaction of the laser-plasma and dipole using an intensified charge-coupled device (ICCD) camera.  The camera viewed along the LAPD central axis through a mirror mounted inside of the LAPD chamber.  Highly temporally-resolved movies were acquired over hundreds or thousands of shots by incrementing the camera delay relative to the laser trigger.

Additionally, swept Langmuir probes were employed to measure $x$-$z$ and $x$-$y$ planes of plasma electron density and temperature near the dipole magnet.  These measurements were carried out in the absence of the laser plasma, and so provide the initial state of the ambient plasma at $t_0$.

\begin{table*}
\small
\centering
\begin{tabular}{p{0.3\textwidth}>{\centering}p{0.1\textwidth}>{\centering}p{0.1\textwidth}>{\centering}p{0.1\textwidth}>{\centering}p{0.1\textwidth}p{0.1\textwidth}<{\centering}}
										&	Run 1			&	Run 2		&	Run 3		&	Run 4		&	Run 5		\\
\hline
\textbf{Background Parameters}				&					&				&				&				&				\\
Dipole magnetic moment $M$					&	0 Am$^2$			&	95 Am$^2$	&	475Am$^2$	&	950 Am$^2$	&	950 Am$^2$	\\
Ion species								&	\multicolumn{4}{c}{H$^{+1}$}											&	---			\\
Density $n_0$								&	\multicolumn{4}{c}{$\sim3\times10^{12}$ cm$^{-3}$}							&	0 cm$^{-3}$	\\
Magnetic field $B_0$	        						&	\multicolumn{4}{c}{300 G}												&	0 G			\\
Electron temperature $T_{e0}$ 					&   	\multicolumn{4}{c}{$\sim5$ eV} 	 										&	---			\\
Electron inertial length $d_{e0}$				&	\multicolumn{4}{c}{0.3 cm}												&	---			\\
Electron gyroperiod $\omega_{ce0}^{-1}$			&	\multicolumn{4}{c}{0.2 ns}												&	---			\\
Ion temperature $T_{i0}$						&   	\multicolumn{4}{c}{$\sim$1 eV} 											&	---			\\
Ion inertial length $d_{i0}$						&	\multicolumn{4}{c}{13.2 cm}											&	---			\\
Ion gyroperiod $\omega_{ci0}^{-1}$				&	\multicolumn{4}{c}{348 ns}												&	---			\\
Alfv\'{e}n speed $v_A$			        			&	\multicolumn{4}{c}{378 km/s}        										&	---			\\
\hline
\textbf{Laser-Driven Parameters}				&					&				&				&				&				\\
Laser energy $E_{laser}$ 		        				&	\multicolumn{5}{c}{20 J}																\\
Plasma speed $v_{l}$						&	\multicolumn{5}{c}{210 km/s}															\\
Ion species								&	\multicolumn{5}{c}{H$^{+1}$, C$^{+1-6}$}													\\
Electron gyroradius $\rho_{e}=v_l/\omega_{ce0}$	&	\multicolumn{4}{c}{40 $\mu$m}											&	---			\\
H ion gyroradius $\rho_{H}=v_l/\omega_{ci0}$		&	\multicolumn{4}{c}{7.3 cm}												&	---			\\
C ion gyroperiod $\omega_{ci0,C}^{-1}$			&	\multicolumn{4}{c}{$(0.7-4.2)\times10^3$ ns}								&	---			\\
C ion gyroradius $\rho_{C}=v_l/\omega_{ci0,C}$	&	\multicolumn{4}{c}{$14.6-87.7$ cm}										&	---			\\
Magnetic cavity speed $v_0$					&	\multicolumn{4}{c}{135 km/s}											&	165 km/s		\\
Magnetic cavity standoff $L_{dia}$				&	11.5 cm			&	13.75 cm		&	15 cm		&	$>15.5$ cm	&	12.25 cm		\\
Magnetopause standoff $L_M$					&	---				&	$<9$ cm		&	13 cm		&	$>15.5$ cm	&	12.25 cm		\\
\hline
\textbf{Dimensionless Parameters}				&					&				&				&				&				\\
Thermal $\beta=8\pi n_0 T_{e0}/B_0^2$			&	\multicolumn{4}{c}{0.01}												&	---			\\
Electron magnetization $\rho_{e0}/d_{i0}$			&	\multicolumn{4}{c}{0.001}												&	---			\\
Ion magnetization $\rho_{i0}/d_{i0}$				&	\multicolumn{4}{c}{0.02}												&	---			\\
Electron collisionality $\omega_{ce0}\tau_{ei}$		&	\multicolumn{4}{c}{$5\times10^{2}$}										&	---			\\
Mach number $M_s=v_l/v_s$					&	\multicolumn{4}{c}{5.3}												&	---			\\
Alfv\'{e}n Mach number $M_A=v_l/v_A$			&	\multicolumn{4}{c}{0.6}		        										&	---			\\ 
Hall parameter $D=L_M/d_{i0}$				&	---				&	$<0.7$		&	1			&	$>1.2$		&	---			\\
\hline
\end{tabular}	
\caption{Summary of experimental runs with typical plasma parameters.  For the laser-driven plasma, parameters are given for the range of C ionization between C$^{+1}$ and C$^{+6}$. The magnetization for species $s$ is calculated with respect to the background gyroradii $\rho_{s0}=v_{th,s}/\omega_{cs0}$, where $v_{th,s}\propto\sqrt{T_s/m_s}$.}
\label{tab:params}
\end{table*}

\section{Results}\label{sec:results}

When measuring the interaction of the laser-driven plasma with the dipole magnetic field, the dipole field evolves too slowly ($\sim$ms) to be measured on the timescales ($\sim$$\mu$s) of the laser-driven plasma.  Instead, the contributions to the total field from the laser-driven plasma interaction and from the dipole magnet were measured in separate runs with the same bdot probe.  Runs with the laser-driven plasma were digitized at 1.25 GHz over a few tens of $\mu$s to record the laser-plasma-dipole interaction.  The same runs without the laser-driven plasma were then digitized at 100 MHz over several ms to cover a full period of the dipole-only field.  The total field during the lifetime of the experiment is then calculated as $\bm{B}_{tot}=\bm{\Delta B} + \bm{B}_{init}$, where $\bm{\Delta B}$ is the field measured during the laser-plasma interaction, and $\bm{B}_{init}=\bm{B}_{dip}+\bm{B}_0$ is the initial unperturbed field due to the slowly-evolving dipole field $\bm{B}_{dip}$ and the uniform background field $\bm{B}_0=B_0 \hat{z}$.

\subsection{Performance of Dipole Magnet}

\begin{figure}[t]
	\centering
	\includegraphics[width=0.45\textwidth]{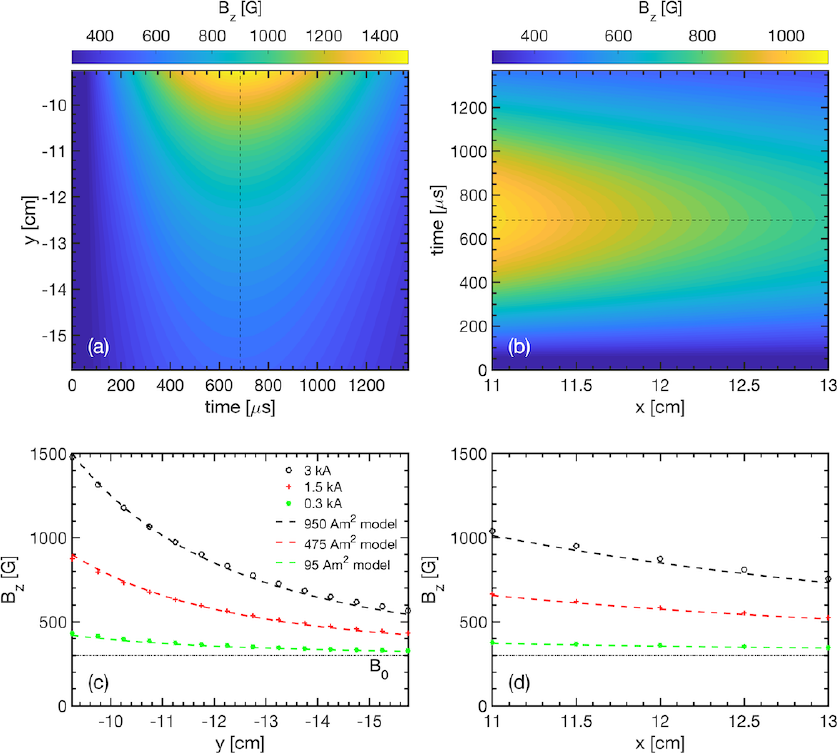}
	\caption{Streak plots of the measured dipole magnetic field (a) along $y$ at $x=0.75$ cm and (b) along $x$ at $y=0$ cm for a 3 kA current.  (c) Comparison of the total magnetic field profiles $B_{z,tot}=B_0+B_{z,dip}$ at the time of peak field ($t\approx685$ $\mu$s) for 3 kA (black), 1.5 kA (red), and 0.3 kA (green) dipole currents.  The field profiles are modeled using the far-field dipole approximation $B_{z,dip}=M/y^3$, where $M$ is the magnetic moment. (d) Similar field profiles and models at the same time in (b).}
	\label{fig:dipole}
\end{figure}

\todo{The performance of the dipole magnet is shown in Fig.~\ref{fig:dipole}(a)-(b) for a dipole coil current of 3 kA. For these measurements, the dipole magnet was embedded in the background field $B_0$ and background plasma, but there was no laser-driven plasma.} While the dipole coil center is nominally located at the center of the LAPD chamber ($\{x,y,z\}=\{0,0,0\}$), measurements indicate that it is slightly offset, with the peak field along $y$ located at $x=0.75$ cm. \todo{At $y=-9$ cm (the closest to the magnet we can measure), the dipole reaches a peak value of $B_{z,dip}\approx1500$ G in $\approx685$ $\mu$s and is constant in magnitude to within 1\% for over 100 $\mu$s (longer than the lifetime of the experiment).}  Fig.~\ref{fig:dipole}(c) shows profiles of the total z-component of the magnetic field $B_{z,tot}=B_0+B_{z,dip}$ along $y$ at $x=0.75$ cm for 3 kA (black), 1.5 kA (red), and 0.3 kA (green) dipole coil currents.  Similar profiles along $x$ at $y=0$ are shown in Fig.~\ref{fig:dipole}(d). The profiles are well-modeled (dashed curves) by the far-field dipole approximation $B_{z,dip}=M/y^3$, where $M$ is the magnetic moment and $y$ is the distance from the dipole center.  For a 3 kA dipole current, the magnetic moment $M_{950}\approx950$ Am$^{2}$.  The moments scale linearly with the current, so that the 1.5 kA and 0.3 kA runs correspond to $M_{475}\approx475$ Am$^{2}$ and $M_{95}\approx95$ Am$^{2}$, respectively.

\subsection{Fast-Gate Imaging}

\begin{figure}
	\centering
	\includegraphics{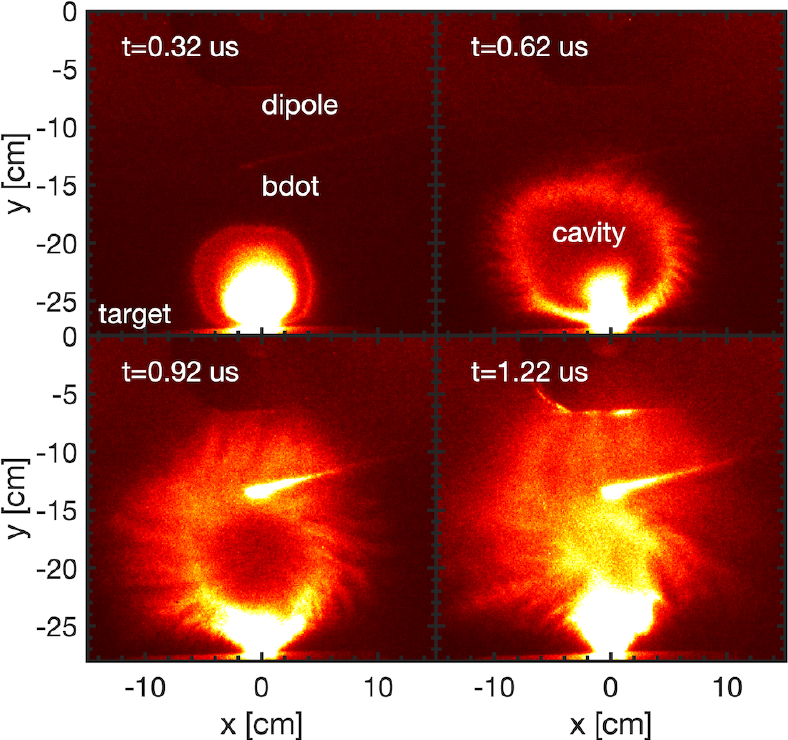}
	\caption{Fast-gate images of plasma optical self-emission for a run with $M=950$ Am$^{2}$.  Each image is gated over 10 ns.  Labeled are the locations of the target and dipole magnet, as well as the bdot probe at the time of the images.  The colorbar is saturated for clarity.}
	\label{fig:pimax}
\end{figure}

To visualize the laser-driven plasma, fast-gate images were acquired.  Example images from a run at $M_{950}$ are shown in Fig.~\ref{fig:pimax}.  Each image is gated over 10 ns and obtained from a different laser shot.  The plastic target is located at the bottom edge of the images, and the dipole magnet center is located at the top.  The bdot probe is also visible near the center of the image.  The laser-ablated plasma is initially approximately spherical in shape, which is then distorted by Rayleigh-Taylor modes in the large-Larmor-radius limit \cite{ripin_large-larmor-radius_1987,huba_theory_1987,collette_structure_2010}.  By $\sim$1 $\mu$s, the plasma has reached the dipole magnet surface.  A cavity is clearly visible in the emission at earlier times, which previous LAPD experiments have shown is closely aligned with a magnetic cavity \cite{schaeffer_a-platform_2018}. This cavity appears to collapse by $t=1.22$ $\mu$s, though material continues to be emitted from the target for several more $\mu$s.  

The laser-driven plasma consists of both H ions and a range of C ionizations \cite{schaeffer_characterization_2016}; however, these images only capture self-emission from the bulk, lower-ionization C components of the laser-driven plasma; the H background plasma, H-component of the laser-driven plasma, and highly-ionized C-component of the laser-driven plasma are not imaged over the wavelengths to which the camera is sensitive.  Since the highly-ionized C or H ions primarily drive the interaction with the dipole magnet (since they have the smallest gyroradii), those effects are not reflected in these images.  Conversely, the bulk C plasma and associated instabilities appear to have little affect on the development of a magnetosphere over the timescales analyzed.

\subsection{Magnetopshere Measurements}


\begin{figure*}[t]
	\centering
	\includegraphics{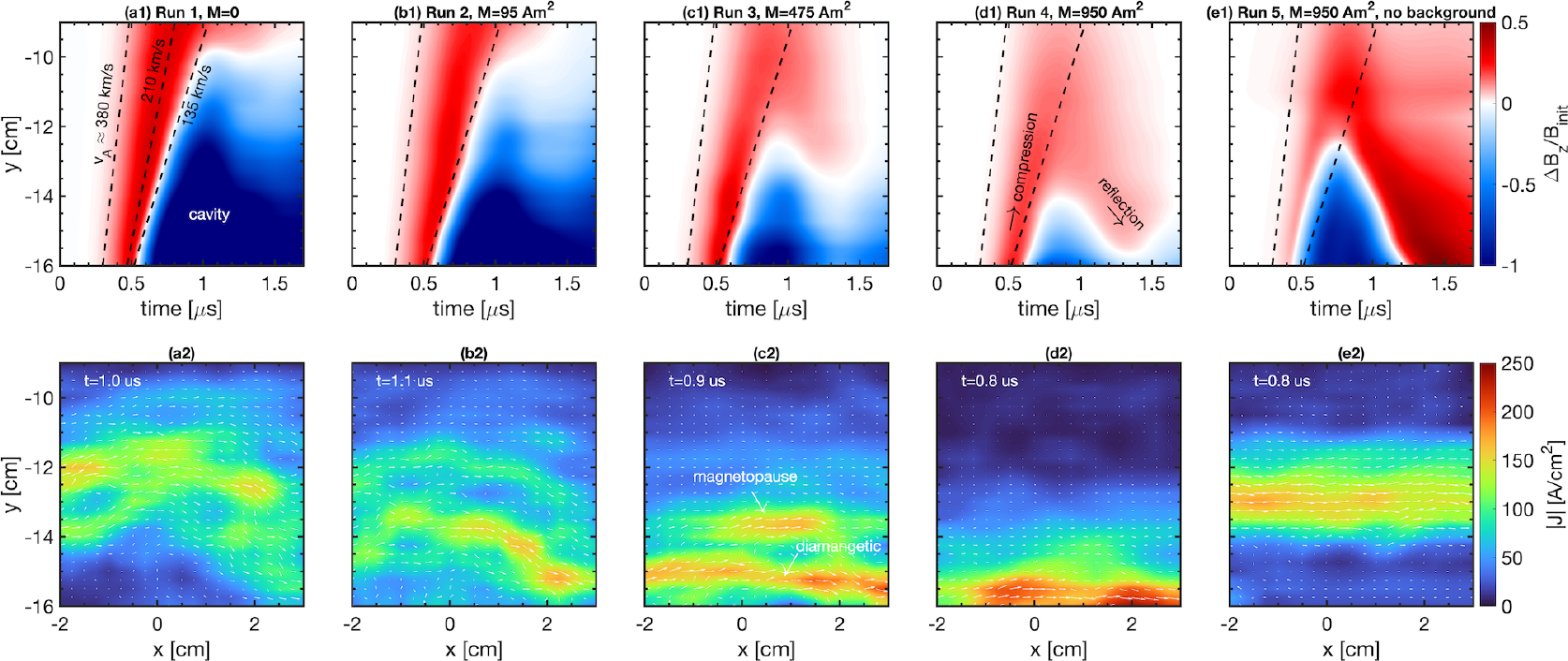}
	\caption{(Top panels) Dayside magnetic field streak plots along $y$ at $\{x,z\}=\{0.75,0\}$ cm for different dipole magnetic moments $M$. In case (e), there is additionally no background plasma or magnetic field $B_0$.  The edge of the dipole magnet is located at $y=-7$ cm.  The colorbars are saturated to make features more clear. (Bottom panels) 2D contour plots of the derived dayside current density in the $x$-$y$ plane, taken at the time of peak current for each $M$. Overplotted is the current density vector field (white arrows).}
	\label{fig:dayside}
\end{figure*}

The laser generates a strongly-driven plasma flow, either directly through the laser-ablated plasma or by accelerating the background plasma \cite{bondarenko_collisionless_2017,bondarenko_laboratory_2017}.  The resulting interaction is shown in Fig. \ref{fig:dayside} for four different dipole moments and for a case with the dipole but without a background plasma or field.  The data consists of 2D $x$-$y$ planes taken on the ``dayside,'' i.e. between the laser target and dipole magnet, that span from $x=-2$ to $x=3$ cm and from $y=-16$ to $y=-9$ cm at $z=0$ (the edge of the dipole magnet extends to $y=-7$ cm).  Each plane was compiled over several thousand laser shots, as described in Sec. \ref{sec:setup}.  The top row consists of streak plots of the relative change in magnetic field $\Delta B_z/B_{init}$ at $x=0.75$ cm (the location of peak dipole field), and the bottom row consists of the corresponding 2D contour plots in the $x$-$y$ plane of current density $J\propto \nabla\times \Delta B_z$ at the time of peak current.  The magnetic field plots were created by averaging over $x=0.25$ to $x=1.25$ cm and then applying a moving average along $y$ with a width of 0.75 cm.  After calculating $J_x$, the current density plots were similarly smoothed.  \todo{A summary of the experimental runs is provided in Table \ref{tab:params}.}

Figure~\ref{fig:dayside}(a1) shows the case with zero dipole moment $M=0$ (the dipole magnet was inserted into the vacuum chamber but not pulsed).  The laser plasma creates a diamagnetic cavity in the background plasma \cite{niemann_dynamics_2013} that completely evacuates the background field ($\Delta B_z/B_{init}\approx-1$).  The peak magnetic compression ($\Delta B_z/B_{init}\approx0.3$) moves at $\sim210$ km/s, which we take as the speed of the laser-driven plasma.  The leading edge of the compression moves at $\sim380$ km/s, comparable to the Alfv\'{e}n speed, while the cavity itself propagates out at approximately 135 km/s.  The speeds are labeled in Fig.~\ref{fig:dayside}(a1), and the leading and cavity speeds are shown as dashed lines for reference in Fig.~\ref{fig:dayside}(b1)-(e1).  The cavity is supported by a strong diamagnetic current that extends across $x$ as seen in Fig.~\ref{fig:dayside}(a2).  After about 1.5 $\mu$s the cavity begins to collapse as the expelled field diffuses back in.   Similar behavior is observed for a low dipole moment $M_{95}$ (see Figs.~\ref{fig:dayside}(b1)-(b2)).

Figure~\ref{fig:dayside}(c1) shows the case with a significantly larger dipole moment $M_{475}$.  The laser-driven cavity and compression are still visible; the initial evolution of the laser-driven plasma is largely unaffected by the additional dipole field due to the $1/y^3$ falloff. However, closer to the dipole magnet, the extra magnetic pressure is able to balance the plasma ram pressure, and the edge of the cavity ($\Delta B_z/B_{init}=0$) only propagates to $y\approx-13$ cm.  The magnetic compression, in turn, penetrates to the edge of the measurement region ($y=-9$ cm), but is then reflected back to $y\approx-13$ cm by the additional dipole magnetic pressure.  The overall magnetic compression between the cavity and dipole now lasts up to 1.5 $\mu$s.  This effect is more pronounced at the strongest dipole moment $M_{950}$ (see Fig.~\ref{fig:dayside}(d1)), where the cavity is even smaller and the magnetic compression is reflected further back towards the target.  Finally, Fig.~\ref{fig:dayside}(e1) shows streak plots for conditions identical to Fig.~\ref{fig:dayside}(d1), but with no background plasma or background magnetic field $B_0$.  The lack of magnetic field near the target ($B_{dip}<50$ G at $y=-27$ cm) leads to a weaker magnetic compression ahead of the cavity, and the cavity is able to propagate closer to the dipole magnet.  There is a clear reflection point around $y\approx-12$ cm, and the reflected compression is significantly stronger and propagates further back towards the target compared to the $M_{950}$ case.

The dipole magnetic pressure leads to additional structure in the current density.  Without the dipole field (Fig.~\ref{fig:dayside}(a1)) or without the background plasma and field (Fig.~\ref{fig:dayside}(e1)), the diamagnetic current propagates out in tandem with the unrestricted cavity.  At $M_{475}$, though, there are two distinct regions of peaked current density (see arrows in Fig.~\ref{fig:dayside}(c2)), which are also seen at $M_{95}$ (Fig.~\ref{fig:dayside}(b2)), though weaker.  The current structures are extended along $x$, consistent with the large plasma plumes created by the laser (see Fig.~\ref{fig:pimax}).  In contrast, the $M_{950}$ case only has one current feature at the edge of the measurement region.

In Figs.~\ref{fig:dayside}(b2)-(c2), the region with the relatively stronger current density closer to the target (farther from the dipole) is the diamagnetic current.  As the cavity expansion is halted, this current reaches a maximum extent and then persists for a few hundred ns before the cavity begins collapsing.  The magnitude of the diamagnetic current density also increases with dipole moment.  Ahead of the diamagnetic current, there is a shorter-lived region of weaker current density at $M_{95}$ and $M_{475}$.  As discussed in Sec.~\ref{sec:discuss}, this current is associated with a location of the magnetopause, i.e. the region of pressure balance between the plasma ram pressure and magnetic pressure.  In the $M_{950}$ case (Fig.~\ref{fig:dayside}(d2)), the current density is even stronger and likely associated with the magnetopause, though it may also overlap with the diamagnetic current.  In all cases, the current structures are of order $d_i$ from the dipole and span electron scales ($\sim d_e$), emphasizing the kinetic nature of this system.

\subsection{Comparison to PIC Simulations}

\begin{figure}
	\centering
	\includegraphics{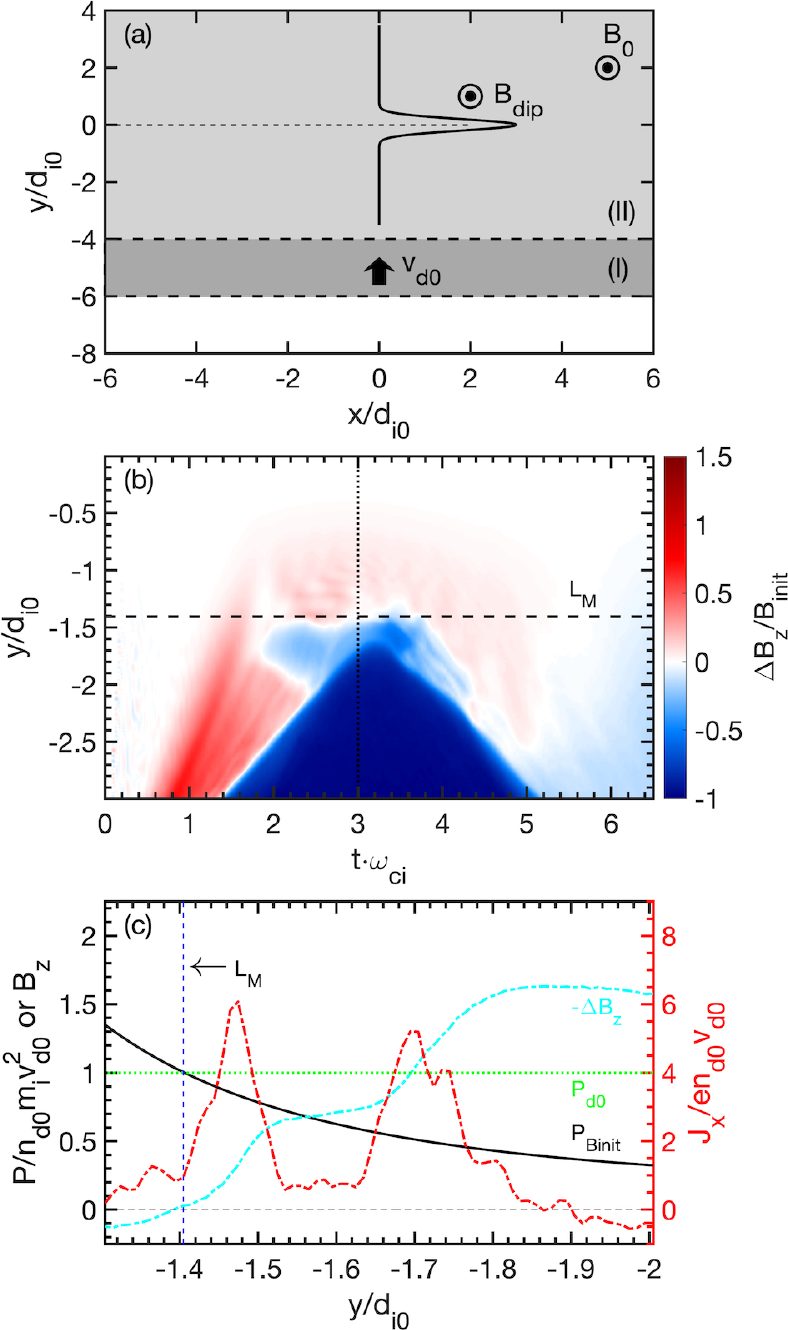}
	\caption{Results from the 2D PIC simulation discussed in the text. (a) Simulation setup.  A uniform driver plasma starts in region (I) with initial velocity $v_{d0}$, and the uniform magnetized background plasma starts in region (II). The dipole field $B_{dip}$ is centered at $\{x,y\}=\{0,0\}$. (b) Streaked contour plot of magnetic field at $x=0$. (c) Profiles of initial magnetic pressure $P_{Binit}$ and initial driver kinetic ram pressure $P_{d0}$, along with the change in magnetic field $-\Delta B_z$ and current density $J_x$ at time $t=3~\omega_{ci}^{-1}$ (dotted line in (b)).  The pressures are defined in the text.}
	\label{fig:sim}
\end{figure}

To further interpret the experimental data, we performed 2D simulations using OSIRIS, a massively parallel, fully relativistic particle-in-cell (PIC) code \cite{fonseca_osiris_2002,fonseca_exploiting_2013}.  Using PIC allows us to accurately resolve the kinetic scales associated with mini-magnetospheres.  In the simulations, a uniform slab representing the laser-driven plasma expands into a uniform background plasma embedded in a combination of a constant magnetic field $B_0$ and a dipole magnetic field.  The simulation setup is shown in Fig.~\ref{fig:sim}(a).  The plasma and field parameters are chosen to be similar to those in the experiment; specifically, the dipole magnetic moment and laser-driven plasma expansion were designed so that the magnetopause standoff was $L_M=1.4~d_{i0}$.  Additionally, to reduce computational resources, a reduced electron-ion mass ratio of $m_e/m_i=100$, increased $v_{d0}/c$ ratio (where $v_{d0}$ is the initial slab plasma speed), and initially cold plasmas were used.  Here, we focus on key results from the simulations for comparison to the data.  Additional information about the simulation setup, and detailed simulation results, are presented in Part II.

Figure~\ref{fig:sim}(b) shows a streaked contour plot of the relative magnetic field $\Delta B_z/B_{init}$ along $y$ at $x=0$ from the simulation.  As in the experiments, the expanding plasma slab drives a diamagnetic cavity and leading magnetic compression.  Similarly, the compression advances past $L_M$ and is then reflected back at later times, while the magnetic cavity is stopped near $L_M$.  Fig.~\ref{fig:sim}(c) shows lineouts from the simulation of $\Delta B_z$ at $t=3~\omega_{ci}^{-1}$ from Fig.~\ref{fig:sim}(b), as well as the current density $J_x$.  Also plotted are the initial total magnetic pressure $P_{Binit}=B_{init}^2/2\mu_0$ and initial driver kinetic ram pressure $P_{d0}=n_{d0}m_dv_{d0}^2$.  As can be seen, there are two peaks in the current density corresponding to the magnetopause current (around $y\approx-1.5 d_{i0}$) and the diamagnetic current (around $y\approx-1.7 d_{i0}$).

The location of these currents is dictated by pressure and energy balances.  By design, the initial driver kinetic pressure is set up to balance the total magnetic pressure, $P_{d0}=P_{Binit}$, at $L_M=1.4~d_{i0}$.  This pressure balance defines the magnetopause and is directly seen in Fig.~\ref{fig:sim}(c), where the magnetopause current peaks slightly behind where $P_{d0}=P_{Binit}$.  Since the laser-driven plasma acts to sweep up and accelerate the background plasma, the furthest extent of the diamagnetic current is dictated by how much of the initial driver energy is used to accelerate background plasma versus expel magnetic field.  In the simulation, approximately 53\% of the initial driver energy goes into the fields by time $t=3~\omega_{ci}^{-1}$.  This energy is used to expel the magnetic field from where the driver starts to the location $L_{dia}$ of the diamagnetic current and can be written \todo{$W_B/W_{d0}=\int_{-4 d_i}^{L_{dia}} P_{Binit} dy/W_{d0}$, where $W_{d0}=P_{d0} L_d$ is the initial driver energy and $L_d$ is the width of the driver.}  For $W_B/W_{d0}=0.53$, this yields $L_{dia}\approx-1.62~d_i$, consistent with the front edge of the diamagnetic current seen in Fig.~\ref{fig:sim}(c).

Based on the detailed simulations presented in Part II, we make here three additional observations that are relevant to the experiments.  First, both the driver and driver-accelerated background ions support the magnetopause.  In the simulations, the background ions, which stream ahead of the bulk of the driver ions, initially establish a magnetopause as a pressure balance between the background ion kinetic ram pressure and the relative magnetic pressure, $P_{bg}=P_{Brel}\equiv(B_{tot}^2-B_0^2)/2\mu_0$.  The relative magnetic pressure is relevant because the background plasma is initially entrained in the background magnetic field, and so the pressure contribution from $B_0$ can be ignored.  Later, another magnetopause is supported by both the driver and background ions where $P_{d0}=P_{Binit}$, since by this time much of the background plasma has been pushed out.  Meanwhile, the diamagnetic current is driven primarily by the driver plasma.

Second, given sufficient energy, the driver plasma will expel magnetic field up to the magnetopause, beyond which the driver plasma does not have sufficient pressure to expand further; in other words, the farthest that the diamagnetic current can be driven is $L_M$.  In the simulations, the driver energy is primarily set by the width of the slab plasma (the initial slab velocity is held constant), with wider slabs equivalent to higher driver energies.  Increasing the slab width will thus push the diamagnetic current closer to the magnetopause location until they merge as a single current structure, which is observed in the simulations presented in Part II.  In the simulation presented here, the driver slab doesn't have enough energy (i.e. energy-constrained) to expel fields up to the magnetopause, resulting in the double current structure observed in Fig.~\ref{fig:sim}(c).  Furthermore, the reflection of the compressed field is also due to the finite driver width; the more energetic the driver, the longer the magnetopause can be maintained before the field is reflected (there is no reflection for an infinite driver).

Lastly, for a given driver energy, simulations observe that the separation between the magnetopause and diamagnetic current decreases with increasing dipole moment $M$.  This can be understood as follows.  On the one hand increasing $M$ will push the magnetopause location $L_M$ farther from the dipole and closer to the location of the diamagnetic current $L_{dia}$.  On the other hand, the increase in magnetic field amplitude means that the the driver depletes a larger fraction of its energy per unit length over the propagated distance, resulting in a smaller diamagnetic cavity.  For a fixed amount of energy into the fields, $L_{dia}$ will increase faster than $L_M$ as $M$ increases, seemingly implying that the separation between the magnetopause and diamagnetic current should increase with dipole moment.  However, the driver also sweeps out a smaller region of background plasma, resulting in less relative energy going into the background plasma and more energy available to expel the fields.  This extra energy is sufficient to compensate for the larger fields and allows the driver to push the diamagnetic current closer to the magnetopause.

\section{Discussion}\label{sec:discuss}

\begin{figure}
	\centering
	\includegraphics{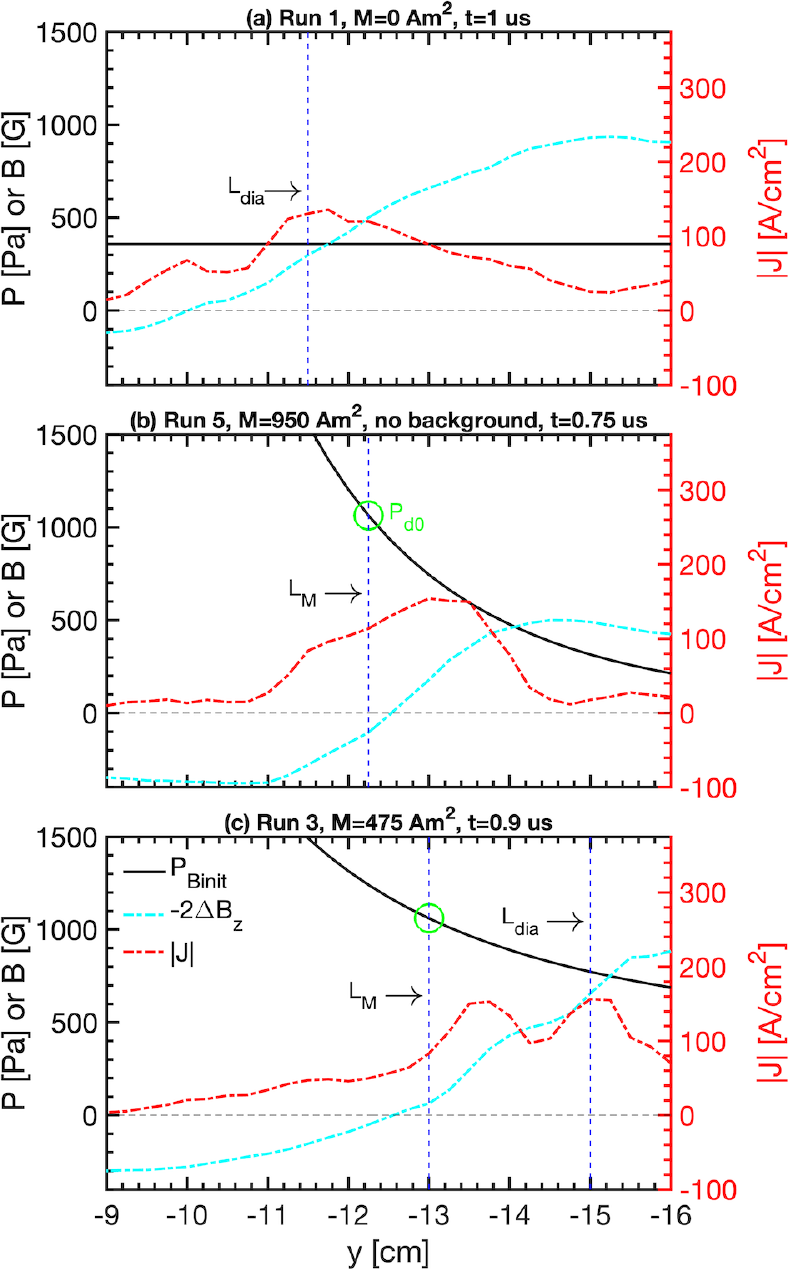}
	\caption{Dayside current density $J$ (red) at $x=0.75$ cm for three $M$ from Figs.~\ref{fig:dayside}(a2), (c2), and (e2). Also plotted are the total initial magnetic pressure $P_{Binit}$ (solid black), change in magnetic field $-\Delta B_z$ (cyan), and the approximate location of the magnetopause $L_M$ or diamagnetic $L_{dia}$ currents (blue).  The green circle indicates the initial driver pressure $P_{d0}$ needed to balance the initial magnetic pressure $P_{Binit}$ at the location of the magnetopause.}
	\label{fig:jx-p}
\end{figure}

Based on the signatures observed in the simulations, in Fig.~\ref{fig:jx-p} we plot lineouts of the current density at $x=0.75$ cm for three cases of $M$ taken from Figs.~\ref{fig:dayside}(a2), (c2), and (e2).  Also plotted are the total initial magnetic pressure $P_{Binit}$ and change in magnetic field $-\Delta B_z$.  With a background plasma and background field, but no dipole field ($M=0$, see Fig.~\ref{fig:jx-p}(a)), the driver pressure is greater than the initial magnetic pressure everywhere ($P_{d0}>P_{Binit}$), and there is no pressure balance, and hence no magnetopause.  Thus, the only current structure created is the diamagnetic current as the driver plasma expands out.  The approximate final position $L_{dia}$ of the diamagnetic current is shown in Fig.~\ref{fig:jx-p}(a).  We can estimate the total initial driver energy per area $\overline{W}_{d0}$ as the sum of the energy needed to expel the field $\overline{W}_B=\int_{L_{tar}}^{L_{dia}} B_0^2/2\mu_0 dy$ and sweep out the background plasma $\overline{W}_{bg}=\int_{L_{tar}}^{L_{dia}} n_0 m_i v_{0}^2 dy$ between the target position $L_{tar}$ and $L_{dia}$, assuming a uniform background plasma.  For the parameters in Table~\ref{tab:params} and $L_{tar}=-27.5$ cm, we find $\overline{W}_{d0}\approx 90$ J/m$^2$.

Figure~\ref{fig:jx-p}(b) shows the case where there is no background plasma or magnetic field, and the driver plasma expands into just the dipole magnetic field.  Here, the driver will expand out, creating a diamagnetic cavity, until it reaches a pressure balance with the total initial magnetic field ($P_{d0}=P_{Binit}$) or runs out of energy.  The energy required to expel the field in Fig.~\ref{fig:jx-p}(b) is only $\overline{W}_{d0}\approx 30$ J/m$^2$.  Since the driver plasma is effectively identical between all runs, this indicates that the driver plasma is not energy-constrained and instead reaches the magnetopause.  Based on the location of the magnetopause $L_M$ in Fig.~\ref{fig:jx-p}(b) (taken where $\Delta B_z\approx0$) and the pressure balance $P_{d0}=P_{Binit}$, we can then estimate $P_{d0}\approx 1050$ Pa.  This is reasonable since it only requires an average drive plasma density $n_{d0}\approx10^{13}$ cm$^{-3}$ and speed $v_{d0}\approx250$ km/s, easily attainable in the experiments.

We expect the same initial driver pressure in Fig.~\ref{fig:jx-p}(c), and the location of $P_{d0}=P_{Binit}$ is shown as the green circle.  As in the simulations, the location of this pressure balance is coincident with the location where $\Delta B_z\approx0$, and slightly behind it ($y\approx-13.5$ cm) we observe a peak in the current density consistent with the magnetopause current.  Also like the simulations, further from the magnetopause ($y\approx-15$ cm) we find the diamagnetic current.  Following a similar calculation as above \todo{and using the location of $L_{dia}$ shown in Fig.~\ref{fig:jx-p}(c)}, we can estimate the energy needed to expel the fields and sweep out background plasma. The total driver energy needed is $\overline{W}_{d0}\approx 90$ J/m$^2$, which is the same as in Fig.~\ref{fig:jx-p}(a).  This implies that the driver plasma does not have enough energy to drive the diamagnetic current all the way to the magnetopause, also consistent with the simulations.  

At $M=95$ Am$^2$, the driver plasma would not reach pressure balance until very close to the dipole, beyond the measurement region.  Since the observed current structures only reach $y\approx-12$ cm (see Fig.~\ref{fig:dayside}(b2)), this indicates that here too the driver plasma runs out of energy before reaching the magnetopause, similar to Fig.~\ref{fig:jx-p}(c).  Taking the current structure at $y\approx-13.75$ cm as the diamagnetic current, the total driver energy needed is $\overline{W}_{d0}\approx 80$ J/m$^2$.  Given the much weaker dipole field, the weak current structure ahead of the diamagnetic current may be a magnetopause driven by background ions rather than driver ions as in the other cases.  A typical background kinetic pressure would be $P_{bg}\sim$100-200 Pa for the values in Table ~\ref{tab:params}, too low to account for the features in the larger $M$ cases but sufficient to balance $P_{Brel}$ near $y\approx-11$ cm.

Finally, it is difficult to conclude anything from the highest moment case $M=950$ Am$^2$ (see Fig.~\ref{fig:dayside}(d2)).  Assuming the same initial driver pressure, the magnetopause would be located at $y\approx-16$ cm, right at the edge of the measurement region.  Assuming the same initial driver energy, we would expect the diamagnetic current to be located around $y\approx-17$ cm (outside the measurement region).  The observed current structure could thus be the magnetopause current.  The diamagnetic current would also be closer to the magnetopause than at lower $M$, consistent with the simulations.

\section{Conclusions}\label{sec:conc}

In this paper, we have presented preliminary results from a new experimental platform to study strongly-driven ion-scale magnetospheres.  The platform -- including background magnetized plasma, target and laser-driven plasma, pulsed dipole magnet, and diagnostics -- can be run at high repetition rate ($\sim1$ Hz), allowing detailed 2D measurements of the magnetic field evolution acquired over thousands of shots.  Data with four different dipole moments ($M=0, 95, 475, \text{and } 950$ Am$^2$) was collected.  In the absence of a dipole field, only the magnetic cavity and associated diamagnetic current from the laser-driven plasma were observed.  \todo{In contrast, for $M>0$ a magnetopause current, in addition to the diamagnetic current, was} observed on kinetic ion and electron scales (i.e. of order $d_i$ and $d_e$), indicating the formation of a mini-magnetosphere.

The experimental results were compared to 2D PIC simulations using the code OSIRIS.  The simulations reproduce the basic magnetic field structures seen in the experiments, including the magnetic compression and cavity formed by the laser-driven plasma, and the reflection of the compression by the dipole pressure.  The simulations confirm that the location of the magnetopause is dictated by the balance between the initial driver kinetic ram pressure and the initial total magnetic field pressure.  However, dynamically the magnetopause current is supported by both the background and laser-driven ions (though the current itself is carried by the electrons) and a complicated time-dependent combination of driver pressure balance and the pressure balance between background ion ram pressure and the relative magnetic pressure (i.e. total magnetic pressure minus the pressure form the constant background field).  The signatures of these pressure balances, derived from the simulations, are also observed in the experiments.  Lastly, the simulations show that as the dipole moment is increased, the location of the magnetopause is pushed further from the dipole.  This results in a shrinking separation between the magnetopause and diamagnetic currents, and even overlapping current structures, features that are observed in the experiments.

While the experiments employed a double cathode setup to create a high density background plasma in the core of the LAPD, the constrained size of the high-density core meant that the laser-driven plasma mostly expanded through a lower density background plasma.  This resulted in a primarily sub-Alfv\'{e}nic ($M_A\approx0.6$) interaction and a Hall parameter of $D\approx1$.  The LAPD has recently implemented a new large-diameter LaB$_6$ cathode that will make most of the background plasma higher density, enabling both super-Alfv\'{e}nic expansions ($D>M_A>1$) and the study of bow shocks.

Future experiments will focus on three main objectives.  First, we will take advantage of the high-repetition-rate platform to expand the 2D planes measured here into 3D cubes to obtain fully 3D magnetic field and current density profiles. Second, in addition to the ``dayside'' we will measure other regions around the dipole, including the ``nightside'' opposite the laser target.  Finally, we will deploy a magnetic field configuration in which the dipole and background fields are anti-aligned in the measurement region (they were aligned in the experiments presented here). This will allow magnetic reconnection in the ``subsolar'' region to be studied and contrasted with the configuration explored in this paper, \todo{in which any reconnection would have been dominantly poleward.}

  
\begin{acknowledgments}

We are grateful to the staff of the BaPSF for their help in carrying out these experiments. The experiments were supported by the DOE under Grants No. DE-SC0008655 and No. DE-SC00016249, by the NSF/DOE Partnership in Basic Plasmas Science and Engineering Award No. PHY-2010248, and by the Defense Threat Reduction Agency and Lawrence Livermore National Security LLC under contract No. B643014.  The simulations were supported by the European Research Council (InPairs ERC-2015-AdG 695088) and FCT (PD/BD/114307/2016 and APPLAuSE PD/00505/2012).  Access to MareNostrum (Barcelona Supercomputing Center, Spain) was awarded through PRACE, and simulations were performed at the IST cluster (Lisbon, Portugal) and at MareNostrum.

\end{acknowledgments}

\bibliography{lapd_minimag_biblib}


\end{document}